\documentclass[prb,
twocolumn,
12pt,
superscriptaddress,showpacs,amsmath,amssymb]{revtex4}
\usepackage{amsfonts}
\usepackage{bm}
\usepackage{verbatim}

\usepackage{graphicx}

\begin{document}

\author{G.E.~Volovik}
\affiliation{Low Temperature Laboratory, Aalto University,  P.O. Box 15100, FI-00076 Aalto, Finland}
\affiliation{Landau Institute for Theoretical Physics, acad. Semyonov av., 1a, 142432,
Chernogolovka, Russia}

\title{Acoustic metric and Planck constants}

\newcommand{\nh}{\slash\!\!\!h}

\date{\today}

\begin{abstract}
Based on Akama-Diakonov (AD) theory of emergent tetrads, it was suggested that one can introduce two Planck constants, 
$\hbar$ and $\nh$, which are the parameters of the corresponding components 
of Minkowski metric, $g^{\mu\nu}_{\rm Mink} = {\rm diag}(-\hbar^2,\nh^2,\nh^2,\nh^2)$. In the AD theory, the interval $ds$ is dimensionless, as a result the metric elements and thus the Planck constants have nonzero dimensions.  The Planck constant $\hbar$ has dimension of time, and the Planck constant $\nh$ has dimension of length. It is natural to compare $\nh$ with the Planck length $l_{\rm P}$. However, this connection remains an open question, because the microscopic (trans-Planckian) physics of the quantum vacuum is not known.
Here we study this question using the effective gravity emerging for sound wave quanta (phonons) in superfluid Bose liquid, where the microscopic physics is known, and the elements of the effective acoustic metric are determined by the parameters of the Bose liquid. Since the acoustic interval is dimensionless, one may introduce the effective "acoustic Planck constants". The acoustic Planck constant $\nh_{\rm ac}$  has dimension of length and is on the order of the interatomic distance. This supports the scenario in which $\nh \sim l_{\rm P}$. We also use the acoustic metric for consideration of dependence of $\hbar$ on the Hubble parameter in expanding Universe.
\end{abstract}
%\pacs{
%}

\maketitle

\newpage
%\tableofcontents

\section{Introduction. Dimensionless physics}

In the Akama-Diakonov theory of quantum gravity, \cite{Akama1978,Diakonov2011,VladimirovDiakonov2012,VladimirovDiakonov2014,ObukhovHehl2012} the fundamental (microscopic) objects are the fermionic fields and the spin connection gauge field. The gravitational field is the secondary object:
gravitational tetrads emerge as the vacuum expectation values of the bilinear combinations  of the operators of the fermionic fields:
\begin{eqnarray}
E^a_\mu=<\hat E^a_\mu> \,,
\nonumber
\\
 \hat E^a_\mu = \frac{1}{2}\left( \Psi^\dagger \gamma^a\partial_\mu  \Psi -  \Psi^\dagger\overleftarrow{\partial_\mu}  \gamma^a\Psi\right) \,.
\label{TetradsFermions}
\end{eqnarray}
Fermionic tetrads in Eq.(\ref{TetradsFermions}), which serve as the order parameter of the symmetry breaking phase transition,  give rise to the metric with lower indices, which is the following bilinear combination of tetrads:
\begin{equation}
g_{\mu\nu}=\eta_{ab}E^a_\mu E^b_\nu \,.
\label{MetricElasticity}
\end{equation}
The important property of this approach to quantum gravity is that the tetrads have dimension either of inverse time or of inverse length,
$[E^a_0]=1/[t]$ and $[E^a_i]=1/[L]$, and the metric elements have dimensions $[g_{00}]=1/[t]^2$, $[g_{ik}]=1/[L]^2$ 
and $[g_{0i}]=1/[Lt]$. With such metric, the interval $ds^2=g_{\mu\nu}dx^\mu dx^\nu$ and the 4-volume element  $d^3x dt \sqrt{-g}$ are dimensionless: $[ds]=1$ and $[d^3x dt \sqrt{-g}]=1$. The same takes place for the other diffeomorphism invariant quantities, such as the cosmological constant $\Lambda$, scalar curvature $R$, scalar field $\Phi$,  particle masses  $M$, etc.\cite{Volovik2021,Volovik2021cont} That is why we cal call the physics in such vacuum as the dimensionless physics.

Dimensionless physics emerges also in some other approaches.  This includes the $BF$-theories of gravity\cite{Schonberg1971,Urbantke1984,Jacobson1991,Obukhov1996,HehlObukhov2003,Friedel2012}  and the model of superplastic vacuum,\cite{KlinkhamerVolovik2019} described in terms of the so-called elasticity tetrads.\cite{Dzyaloshinskii1980,NissinenVolovik2019,Nissinen2020,Nissinen2020a,Nissinen2020b,Burkov2021}
Some of these features emerge also for acoustic metric,\cite{Unruh1981,Barcelo2011} see Refs.\cite{Volovik2021,Volovik2021cont}. Here we shall use the properties of the acoustic metric  for the discussion of the problems related to the Planck constants.

\section{Dimensionless physics and Planck constants}

\subsection{Planck constants as elements of Minkowski metric}

The important consequence of the dimensionless physics is that in Minkowski spacetime the Planck constant appears as the element of Minkowski metric.\cite{Volovik2009,Volovik2023a} However, the time and space elements of Minkowski metric require two different Planck constants.\cite{Volovik2023b} 
For example, for massive particles in the non-relativistic limit the wave equation has the following form:
\begin{equation}
i\sqrt{-g^{00}_{\rm Mink}}\partial_t \psi =- \frac{1}{2M}g^{ik}_{\rm Mink}  \nabla_i \nabla_k\psi \,.
\label{SchrodingerEq1}
\end{equation}
Here the mass $M$ is the rest energy, which is diffeomorphism invariant and thus is dimensionless in the AD approach.
Eq.(\ref{SchrodingerEq1}) looks as the Schr\"odinger wave equation:
\begin{equation}
i\hbar \partial_t \psi =-\frac{\nh^2}{2M} \nabla^2\psi \,,
\label{SchrodingerEq}
\end{equation}
which contains two Planck constants:
\begin{equation}
\sqrt{-g^{00}_{\rm Mink}} =\hbar\,\,,\,\, g^{ik}_{\rm Mink} =\nh^2 \delta^{ik}\,.
\label{MinkowskiMetric}
\end{equation}

The Planck constants  $\hbar$ and $\nh$ enter correspondingly the time derivative and space derivative terms in Schr\"odinger equation, and have different dimensions. The Planck constant $\hbar$ has dimension of time, $[\hbar]=[t]$, while the second Planck constant $\nh$ has dimension of length, $[\nh]=[L]$.

\subsection{Length dimension of Planck constant and Planck length}

In dimensionless physics the Newton constant has the dimension of length, $[G]=[L]$, i.e. the same dimension as the spacelike Planck constant $\nh$. There are two quantities, which can be constructed by combination of the Newton constant $G$ and $\nh$. One of them is the Planck mass $M_{\rm P}=\sqrt{\nh/G}$. It enters the Einstein equations as the diffeomorphism invariant quantity, and thus is dimensionless, 
$[M_{\rm P}]^2=[\nh]/[G]=1$. Another quantity is the Planck length 
\begin{equation}
l_{\rm P}=\frac{\nh}{M_{\rm P}}=\sqrt{\nh G}\,,
\label{PlanckLength}
\end{equation}
 with the natural dimension of length, $[l_{\rm P}]=[L]$. Since the Planck length $l_{\rm P}$ and Planck constant $\nh$ have the same dimension, the natural suggestion arises: maybe    they are the equivalent  quantities,  $\nh\equiv l_{\rm P}$.
This connection remains an open question, because the microscopic (trans-Planckian) physics of the quantum vacuum is not known.
However we can study this problem using the system where the microscopic physics is well known. Here we consider an example of the acoustic metric,\cite{Unruh1981,Barcelo2011}  the effective metric emerging for sound wave quanta (phonons) in superfluid Bose liquid,\cite{Volovik2003} such as liquid $^4$He. The microscopic physics in liquid helium is known: it is atomic physics. 

\section{Acoustic Planck constants}

 \subsection{Acoustic metric}
 \label{AcousticMetric}

In acoustic gravity of superfluid $^4$He, the effective acoustic metric and thus the acoustic Planck constants, $\hbar_{\rm ac}$ and $\nh_{\rm ac}$, can be expressed in terms of the parameters of this liquid. We consider the liquid helium in the state at zero temperature and at zero external pressure. This self-sustained liquid represents the  "superfluid quantum vacuum" -- the analogue of the self-sustained vacuum.\cite{KlinkhamerVolovik2008a,KlinkhamerVolovik2008}

If the chosen variable in the superfluid hydrodynamics is the phase $\Phi$ of the Bose condensate, which is dimensionless, then the action for phonons propagating in moving liquid is:\cite{Volovik2003}
\begin{eqnarray}
S_{\rm ph}=\frac{S}{\hbar}=
\nonumber
\\
=\frac{m}{2\hbar} \int d^3x dt \, n \left( (\nabla \Phi)^2  -
 \frac{1}{s^2} (\dot \Phi -{\bf v} \cdot \nabla \Phi)^2\right)=
 \nonumber
 \\
 =\frac{1}{2}\int d^4 x\,\sqrt{-g} \, g^{\mu\nu} \nabla_\mu \Phi \nabla_\nu \Phi \,.
\label{PhononAction}
\end{eqnarray}
Here $n$ is the density of bosonic particles; $m$ is the particle mass; $s$ is the speed of sound; ${\bf v}$ is the superfluid velocity (the velocity of the "superfluid vacuum"): it is the shift function in the Arnowitt-Deser-Misner (ADM)  approach. Finally, $\hbar$ is the conventional Planck constant describing the microscopic physics. All the parameters are considered in the conventional units, i.e. without application of dimensionless physics. The dimensionless physics will emerge for phonons, and this is the reason why we introduced the phonon action $S_{\rm ph}$ as the action divided by $\hbar$.

The corresponding acoustic interval is
\begin{eqnarray}
ds^2=g_{\mu\nu}dx^\mu dx^\nu=
\nonumber
\\
\frac{\hbar n}{ms}[-s^2 dt^2 + (dx^i -v^idt)\delta_{ij}  (dx^j -v^jdt)]\,.
\label{AcousticInterval}
\end{eqnarray}
The analog of the Minkowski metric corresponds to the zero value of the shift function, ${\bf v}=0$, and thus $g^{0i}=0$. Then
the effective acoustic Minkowski metric experienced by the propagating phonons is:
\begin{eqnarray}
g_{00}=\frac{\hbar n s}{m}   \,\,,\,\, g_{ik}=\frac{\hbar n}{ms}  \delta_{ik} \,\,,\,\,
\sqrt{-g}= \frac{\hbar^2n^2}{m^2s} \,,
\label{Effective_metric}
\end{eqnarray}
with dimensions 
\begin{eqnarray}
[g_{00}]=\frac{1}{[t]^2}   \,\,,\,\, [g_{ik}]=\frac{1}{[L]^2} \,\,,\,\,
[\sqrt{-g}]= \frac{1}{[t][L]^3}  \,.
\label{EffectiveMetricDimension}
\end{eqnarray}
The acoustic interval (\ref{AcousticInterval}) is dimensionless, $[ds]=1$. This demonstrates that  the interval $ds$ describes the dynamics of phonons in the superfluid "vacuum", rather than the distances and time intervals.
The same is valid for the interval in general relativity, where it describes the dynamics of a point particle in the relativistic quantum vacuum.

 \subsection{Acoustic quantum mechanics}
\label{AcousticQM}

The effective quantum mechanics and the corresponding Planck constants can be considered using the equation for massive phonons (the quasi-Goldstone or pseudo-Goldstone modes). The massive pseudo-Goldstone phonon exists, for example, in magnon BEC, where the phonon mass is created by application of the RF field.\cite{NissinenVolovik2017,BunkovVolovik2013} If one neglects the magnetic anisotropy, then the effective metric in magnon BEC is given by Eq.(\ref{Effective_metric}), where now 
$n$ is the density of excited magnons, $s$ is the velocity of spin waves and $m$ is magnon mass.\cite{BunkovVolovik2013} 
The massive phonons obey the Klein-Gordon equation for the real scalar field, which is obtained from the action (\ref{PhononAction}) with the added mass term:
\begin{eqnarray}
S_{\rm ph} =\frac{1}{2}\int d^4 x\,\sqrt{-g} \left(g^{\mu\nu} \nabla_\mu \Phi \nabla_\nu \Phi + M^2 \Phi^2\right)\,,
\label{PhononAction2}
\end{eqnarray}
where $M$ is the corresponding  mass of phonon (in magnon BEC, $M^2 \propto H_{\rm rf}$, where $H_{\rm rf}$ is the radio frequency magnetic field). It is important that with the effective acoustic metric (\ref{Effective_metric}) this mass $M$ is dimensionless. This follows from the dimensionless 4-volume $d^3x dt \sqrt{-g}$, see Eq.(\ref{EffectiveMetricDimension}).

For massive phonon there is the  relation between the phonon frequency and its mass, $M=\sqrt{-g^{00}} \omega(k=0)$. This suggests that $\sqrt{-g^{00}}$ plays the role of the Planck constant, the acoustic Planck constant, $\hbar_{\rm ac}=\sqrt{-g^{00}}$. 
This acoustic Planck constant enters  the corresponding Schr\"odinger  equation for phonons, which can be obtained by introducing the phonon wave function $\psi$:
\begin{eqnarray}
\Phi({\bf r},t) = \frac{1}{\sqrt{M}}\exp\left(-i Mt /\sqrt{-g^{00}}\right)\psi({\bf r},t) 
\nonumber
\\
+  \frac{1}{\sqrt{M}}\exp\left(i Mt /\sqrt{-g^{00}}\right)\psi^*({\bf r},t)  \,.
\label{eq:PhiPsiPhonon}
\end{eqnarray}
While the wave function $\Phi$ is real, the wave function $\psi$ is complex. There is no doubling of the degrees of freedom, since $\psi$ combines the solutions with positive and negative frequencies.

For phonons with positive mass $M$, the Klein-Gordon equation reduces in the long wavelength limit to the following equation:
\begin{eqnarray}
i\sqrt{-g^{00}}\partial_t \psi =-\frac{g^{ik}}{2M}\nabla_i\nabla_k \psi   \,.
\label{eq:SchroedingerEq}
\end{eqnarray}
This looks as the Schr\"odinger  equation
\begin{equation}
i\hbar_{\rm ac} \partial_t \psi =-\frac{\nh_{\rm ac}^2}{2M} \nabla^2\psi \,,
\label{SchrodingerEqAc}
\end{equation}
where instead of the conventional Planck constants one has the effective acoustic Planck constants, 
$\hbar_{\rm ac}$ and $\nh_{\rm ac}$:
\begin{eqnarray}
\hbar_{\rm ac}^2=g^{00}=\frac{m}{\hbar n s}  \,\,,\,\, \nh_{\rm ac}^2=\hbar_{\rm ac}^2s^2 =\frac{ms}{\hbar n}  \,.
\label{Effective_hbar}
\end{eqnarray}

\subsection{Tolman temperature for phonons and temperature of the background  liquid}
\label{Tolman}

Thermal distribution of massless (gapless) phonons in inhomogeneous liquid follows the Tolman law:\cite{VolovikZelnikov2003}
\begin{equation}
T({\bf r})=\frac{T_0}{ \sqrt{-g_{00}({\bf r})} } \,.
\label{TTolman}
\end{equation}
Here $T({\bf r})$ is the local temperature measured by the "inhabitants of the world of phonons", and 
$T_0$ is the constant parameter -- the Tolman temperature, which has dimension of frequency, 
$[T_0]=1/[t]$.  The thermodynamic  free energy of phonons in thermal equilibrium is:
\begin{equation}
F=-\frac {\pi^2}{90} \,{T_0^4} \int d^3 x \,\frac{\sqrt{-g({\bf r})}} {g_{00}^2({\bf r})}=
-\frac {\pi^2}{90} \,{T_0^4} \int \,\frac{ d^3 x} {s^3({\bf r})}\,.
\label{FreeEenergy}
\end{equation}
From Eq.(\ref{FreeEenergy}) it follows that the Tolman temperature for phonons $T_0$ coincides with the temperature of the background liquid, when it is expressed in terms of frequency, 
$T_0=T_{\rm liquid}/\hbar$. The temperature of the inhomogeneous liquid is constant in thermal equilibrium, 
 $T_{\rm liquid}={\rm const}$.  
 
 All this demonstrates the connection  between the two worlds with different Planck constants: the microscopic world of atomic physics of the liquid with the Planck constant $\hbar$ and the macroscopic world of phonons with their effective acoustic Planck constant $\hbar_{\rm ac}$.
 The corresponding temperatures are:  $T_{\rm liquid}=\hbar T_0$ and $T_{\rm phonon}=\hbar_{\rm ac} T_0$. These temperatures obey the Tolman law,  $\hbar/T_{\rm liquid}=\hbar_{\rm ac}/T_{\rm phonon}$. This is similar to the Tolman law describing the thermal contact between two different Minkowski quantum vacua with two different Planck constants.\cite{Volovik2023a} But now it describes the "thermal contact" between microscopic and macroscopic physics.
 
 In both worlds, microscopic and macroscopic, the temperature of the Hawking radiation from the acoustic horizon is given by 
 \begin{equation}
T_0= \frac{v'}{2\pi} \,.
\label{HawkingPhonon}
\end{equation}
Here $v'$ is the derivative of the shift function at the acoustic horizon,\cite{Unruh1981,Barcelo2011} while the temperature is measured far away from the horizon, where acoustic metric corresponds to the ground state of the liquid. This temperature has the dimension of frequency, $[T_0]=[\omega]$, and corresponds to the thermal factor $\exp{(-\omega/T_0})$, where $\omega$ is the frequency of the radiated phonon (here we assume the limit of large frequencies, $\omega \gg T_0$). 

The same equation (\ref{HawkingPhonon}) is valid for the Hawking radiation of photons from the horizon of the Schwarzschild black hole. The Tolman temperature $T_0= \frac{v'}{2\pi} =\frac{c}{4\pi r_h}$, where $r_h$ is the horizon radius, also has dimension of frequency, while the temperature $T=\frac{\nh}{4\pi r_h}$ is dimensionless, since $[\nh]=[r_h]=[L]$.

\subsection{Length dimension of acoustic Planck constant and the UV length scale}
\label{AcousticMetricLength}

From Eq.(\ref{Effective_hbar}) it follows that the effective acoustic Planck constant $\nh_{\rm ac}$ has dimension of length, $[\nh_{\rm ac}]=[L]$, and is on the order of the corresponding ultraviolet (UV) length scale, which in quantum liquid is the interatomic distance $a=n^{-1/3}$. Let us consider this on example of liquid helium. It is the self-sustained system, which can live in the absence of external pressure.  At $T=0$ it serves as an analogue of quantum vacuum. There are three UV (atomic) energy scales, which in the strongly correlated and strongly interacting liquid helium have the same orders of magnitude, $E_{{\rm UV}1}\sim E_{{\rm UV}2}\sim E_{{\rm UV}3}$:
\begin{eqnarray}
E_{{\rm UV}1}=ms^2\,,
\label{EffectivePlanck1}
\\
E_{{\rm UV}2}=\frac{\hbar s}{a}\,,
\label{EffectivePlanck2}
\\
E_{{\rm UV}3} = \frac{\hbar^2}{ma^2}\,.
\label{EffectivePlanck3}
\end{eqnarray}
They enter the acoustic Planck constant $\hbar_{\rm ac}$, which has dimension of time: 
\begin{eqnarray}
\hbar_{\rm ac}^2 = \frac{\hbar^2}{E_{{\rm UV}2} E_{{\rm UV}3}} \equiv \frac{1}{\omega_{{\rm UV}2} \omega_{{\rm UV}3}} \,\,,\,\,  [\hbar_{\rm ac}]= [t]\,,
\label{EffectivePlanck4}
\end{eqnarray}
where $\omega_{\rm UV}$ are the corresponding frequencies.

Let us consider for simplicity the idealized quantum liquid, in which all three UV energy scales are the same, $E_{{\rm UV}1}= E_{{\rm UV}2}= E_{{\rm UV}3}$. This implies the following relation between the parameters of the quantum liquid: $msa=\hbar$. Then one has $\hbar_{\rm ac}=a/s$ and $\nh_{\rm ac}=\hbar_{\rm ac} s=a$. This means that in the acoustic analog of quantum vacuum, the length of the space-like Planck constant  $\nh_{\rm ac}$ coincides with the UV length scale --  the interatomic distance $a$. Correspondingly, the time-like acoustic Planck constant coincides with the UV time scale, $\hbar_{\rm ac}=a/s=1/\omega_{\rm UV}$.

In liquid helium there is no direct analog of the Newton constant $G$, and thus there is no direct connection between the UV energy scale and the Planck scale.
Nevertheless, there is some term in the energy of the flowing liquid, which can be interpreted as coming from the curvature of acoustic  spacetime.\cite{VolovikZelnikov2003} 
If using this term, one estimates the effective Newton constant, one obtains that the corresponding Planck energy scale is on the order of the UV energy scale, $E_{\rm  P} \sim E_{\rm UV}$, and the acoustic Planck length is on the order of the interatomic distance  $a$. Then the acoustic Planck mass, being expressed in terms of $\hbar_{\rm ac}$, is on the order of unity,  $M^{\rm P}_{\rm ac}\sim  \hbar_{\rm ac}\omega_{\rm UV} \equiv  \hbar_{\rm ac}\omega^{\rm P}_{\rm ac}=1$.  

Extending this observation to relativistic physics, one may suggest that in principle it is possible that the relativistic  quantum vacuum has the similar property, i.e.  the space-like Planck constant is equal to (or on the order of) the Planck length, $\nh = l_{\rm P}$.  Then the UV energy scale coincides with the Planck energy scale, $M_{\rm UV}=M_{\rm P}=1$.  
On the other hand it is not excluded that the UV energy scale can be much larger than the Planck energy, $M_{\rm UV}>>M_{\rm P}$, see e.g. Refs.\cite{Bjorken2001,KlinkhamerVolovik2005}.

\subsection{Variation of Planck constants in expanding Universe}
\label{AcousticDS}

Applying the dimensionless physics to the de Sitter (dS) expansion, one obtains that the de Sitter vacuum contains three parameters, 
$\hbar$, $\nh$ and the Hubble parameter $H$.  In the Paineve-Gullstrand form, the interval in the dS space has the form of Eq.(\ref{AcousticInterval}) with the shift function $v=Hr$:
\begin{equation}
ds^2 =-\frac{1}{\hbar^2}dt^2  +\frac{1}{\nh^2}\left( (dr -Hrdt)^2 + r^2 d\Omega^2 \right)\,.
\label{dSds}
\end{equation}
The Hubble parameter has dimension of frequency, $[H]=1/[t]$. The Tolman temperature of the Hawking radiation, $T_0=v'/2\pi=H/2\pi$, has also dimension of frequency, which meets the thermal factor $\exp{(-\omega/T_0})$. The corresponding Gibbons-Hawking temperature $T_{\rm GH}=\hbar H/2\pi$ is dimensionless.

At $r=0$ the metric is Minkowski, 
but in principle, it is not excluded that the Planck constants $\hbar$ and $\nh$  may depend on the Hubble parameter $H$. Let us consider this possibility, using again the dimensionless physics of superfluid $^4$He.
 
In the previous sections, we considered liquid helium in its self-sustained vacuum state, i.e. in the ground state at $T=0$ and in the absence of external pressure, $P=0$. This vacuum state has the well determined fixed values for acoustic Planck constants in Eq.(\ref{Effective_hbar}).  In applied pressure $P\neq 0$, the parameters of the quantum vacuum $s$ and $a$ deviate from their vacuum values. For  pressure small compared with the UV scale, $P \ll nms^2$, the relative change of these parameters is small, $\Delta s/s \sim \Delta a/a \sim P/nms^2 \ll 1$.
The same  is valid for the relative change  of the acoustic Planck constants:
\begin{equation}
\frac{\Delta \nh_{\rm ac}}{\nh_{\rm ac}} \sim \frac{\Delta \hbar_{\rm ac}}{\hbar_{\rm ac}} \sim \frac{P}{nms^2}\ll1\,.
\label{VariationAc}
\end{equation}

In relativistic quantum vacuum, the non-zero vacuum pressure gives rise to the de Sitter expansion, with $H^2$ being proportional to pressure $P$. Then, applying Eq.(\ref{VariationAc}) with the corresponding Planck scale parameters, one may suggest that the Planck constants acquire the following corrections in the expanding Universe:
\begin{equation}
\frac{\Delta \nh}{\nh} \sim \frac{\Delta \hbar}{\hbar} \sim\hbar^2H^2 \sim T_{\rm GH}^2\ll1\,.
\label{Variation}
\end{equation}
In the history of the expansion of the Universe such corrections are typically small, and thus the Planck constants, and the corresponding speed of light $c= \nh/\hbar$ are practically constant, as distinct from many theories of the varying speed of light.\cite{Barrow1999,Albrecht1999}

\section{Conclusion}

The important consequence of the composite tetrads approach to quantum gravity is the "dimensionless physics": all the diffeomorphism invariant quantities are dimensionless for any dimension of spacetime. These include the action $S$, interval $s$, cosmological constant $\Lambda$, Hawking temperature $T_H$, scalar curvature $R$, scalar field $\Phi$, 
Planck mass $M_{\rm P}$, masses $M$ of particles and fields, etc.  

The Akama-Diakonov theory also suggests that the Planck constants $\hbar$ and $\nh$ are the elements of Minkowski vacuum. These parameters are not diffeomorphism invariant and thus are not dimensionless, with dimension of time and length correspondingly. Using as an example the effective acoustic Planck constants emerging in liquid helium, where the trans-Planckian (atomic) physics is known, we considered two problems related to the Planck constants $\hbar$ and $\nh$ in relativistic vacuum.

(i) The connection between the length of the Planck constant $\nh$ and the Planck length $l_{\rm P}=\sqrt{\nh G}$. In acoustic gravity for pnonons in liquid helium, the length of the acoustic Planck constant $\nh_{\rm ac}$ is on the order of the interatomic distance $a$. This suggests that in quantum vacuum, the spacelike Planck constant is on the order of the Planck length $l_{\rm P}$. If this is so, then the Planck mass  $M_{\rm P}=\sqrt{\nh /G}$, which enters the Einstein–Hilbert action $S = \frac{M_{\rm P}^2}{16\pi}  \int d^4x\,\sqrt{-g} \,R $ and which is dimensionless as all the masses in the Akama-Diakonov quantum gravity, is on the order of unity, $M_{\rm P}\sim 1$. That is why the Planck mass becomes the natural choice for the unit of mass.

 (ii) The possible variation of the Planck constant in expanding Universe. The application of external pressure to liquid helium changes the values of the acoustic Planck constants according to Eq.(\ref{VariationAc}). In relativistic quantum vacuum, the vacuum pressure gives rise to the de Sitter expansion. This suggests that in the expanding Universe the Planck constants vary according to Eq.(\ref{Variation}).
 
 {\bf Acknowledgements}. This work has been supported by Academy of Finland (grant 332964).

\end{document}